\documentclass[12pt,twoside]{article}
\usepackage[mathscr]{eucal}
\usepackage{amsmath,amsfonts,amssymb,amsthm,mathabx,empheq}
\usepackage{times}
\usepackage{pdfsync}
\usepackage{cite}
\usepackage{url}
\usepackage{hyperref}
\usepackage{tensor}
\usepackage{color}
\usepackage{multicol}
\usepackage{bbold}

\advance\textheight\topskip
\allowdisplaybreaks[1]

\newcommand\td{\text{d}}
\newcommand\cO{{\cal O}}
\newcommand{\p}{\partial}

\newcommand{\be}{\begin{equation}}
\newcommand{\ee}{\end{equation}}
\newcommand{\bea}{\begin{eqnarray}}
\newcommand{\eea}{\end{eqnarray}}

\def\bz{\bar z}

\def\n{\nabla}

\newcommand*\xbar[1]{%
  \hbox{%
    \vbox{%
      \hrule height 0.5pt 
      \kern0.3ex
      \hbox{%
        \kern-0.0em
        \ensuremath{#1}%
        \kern-0.0em
      }%
    }%
  }%
}

\DeclareFontFamily{OT1}{rsfs}{} \DeclareFontShape{OT1}{rsfs}{m}{n}{
<-7> rsfs5 <7-10> rsfs7 <10-> rsfs10}{}
\DeclareMathAlphabet{\mycal}{OT1}{rsfs}{m}{n}

\begin{document}
\title{Note on the asymptotic structure of Kerr-Schild form}

\author{Pujian Mao and Weicheng Zhao}

\date{}

\def\mytitle{Note on the asymptotic structure of Kerr-Schild form}

\addtolength{\headsep}{4pt}

\begin{centering}

  \vspace{1cm}

  \textbf{\Large{\mytitle}}

  \vspace{1.5cm}

  {\large Pujian Mao and Weicheng Zhao}

\vspace{.5cm}

\vspace{.5cm}
\begin{minipage}{.9\textwidth}\small \it  \begin{center}
     Center for Joint Quantum Studies and Department of Physics,\\
     School of Science, Tianjin University, 135 Yaguan Road, Tianjin 300350, China
 \end{center}
\end{minipage}

\end{centering}


\vspace{1cm}

\begin{center}
\begin{minipage}{.9\textwidth}
  \textsc{Abstract}. The Kerr-Schild form provides a natural way of realizing the classical double copy that relates exact solutions in general relativity to exact solutions in gauge theory. In this paper, we examine the asymptotic structure of Kerr-Schild form. In Newman-Unti gauge, we find a generic solution space satisfying the Kerr-Schild form in series expansion around null infinity. The news function in the solution space is chiral and can not lead to a mass loss formula. A class of asymptotically flat complex pp-wave solutions in closed form is obtained from the solution space.
 \end{minipage}
\end{center}
\thispagestyle{empty}


\section{Introduction}

The double copy structure \cite{Bern:2008qj,Bern:2010ue} reveals a remarkable relation between gravity and gauge theory in the context of scattering amplitudes. Apart from the fruitful results and applications (see, e.g., \cite{Bern:2019prr} for a review and references therein), it is still somewhat mysterious that what is the interpretation of double copy from the field theory side, in particular at the classical level. In a pioneering paper \cite{Monteiro:2014cda}, an exact classical double copy was demonstrated for a particular class of solutions in gravity that admits a Kerr-Schild form. Shortly, this type of classical double copy has been extensively investigated and developed \cite{Luna:2015paa,Luna:2016due,White:2016jzc,Bahjat-Abbas:2017htu,Carrillo-Gonzalez:2017iyj,Goldberger:2017ogt,Li:2018qap,Lee:2018gxc,Berman:2018hwd,Gurses:2018ckx,
Luna:2018dpt,CarrilloGonzalez:2019gof,Cho:2019ype,PV:2019uuv,Bautista:2019evw,Kim:2019jwm,Bah:2019sda,Arkani-Hamed:2019ymq,Lescano:2020nve,Gumus:2020hbb,Elor:2020nqe,Easson:2020esh,
Godazgar:2020zbv,Berman:2020xvs,Prabhu:2020avf,White:2020sfn,Monteiro:2020plf,Lescano:2021ooe,Campiglia:2021srh,Alkac:2021bav,Chacon:2021wbr,Alkac:2021seh,Angus:2021zhy,Chacon:2021hfe,
Diaz-Jaramillo:2021wtl,Gonzo:2021drq,Cho:2021nim,Godazgar:2021iae,Adamo:2021dfg}.

The success of Kerr-Schild double copy is temporarily subject to algebraically special solutions.
One of the remarkable features of the Kerr-Schild metric (see, e.g., \cite{Stephani:2003tm} for detailed introduction) is the fact that the null vector and the scalar field in the decomposition can be interpreted as perturbations on the flat background. Such a property is very crucial in connecting solutions of gravitational theory to gauge theory. Given the fact that the numbers of known exact solutions of gravity are indeed limited, in particular the ones admitting a Kerr-Schild form in four dimensional Einstein gravity, the connection between gravity and gauge theory in the Kerr-Schild double copy formalism is very restricted.
However, if one considers solutions of Einstein equation in a perturbative way, namely, given in series expansion, there can be a solution space including infinite amount of solutions. In the seminal work by Bondi et. al. \cite{Bondi:1962px}, it is shown that when expanding the metric
fields in inverse powers of a radius coordinate in asymptotic region, the Einstein equation is formulated as a characteristic initial value problem. In this framework, the gravitational radiation is characterized by the news functions. Once reasonable news functions and initial data are given, a solution of Einstein equation is determined. Naturally one would wonder that if the asymptotic framework can allow us to extend the application of classical double copy.

Indeed, such idea has already been implemented in literature recently. In \cite{Godazgar:2021iae}, the authors consider the rotating STU
supergravity black holes \cite{Chong:2004na}, which are only algebraically special asymptotically. The classical double copy is confirmed at the leading order in the asymptotic expansion. In \cite{Adamo:2021dfg}, a precise prescription is given to connect asymptomatically flat gauge and gravitational fields. In this paper, we start from an alternative point of view. We adopt the Kerr-Schild form in the asymptotic framework and study the solution space admitting Kerr-Schild form in asymptotic expansion around null infinity in four dimensional spacetime.
This computation is also motivated by the triangle relation \cite{Strominger:2017zoo} where the scattering amplitudes in the low energy limit, i.e., soft theorems, are connected to asymptotic symmetries and memory effects. Clearly the soft theorems obey a double copy structure that the soft graviton theorem is double the soft photon one. While the study of asymptotic symmetry and memories are mostly done in the region around null infinity. The Newman-Unti (NU) gauge \cite{Newman:1962cia} is adopted to derive the solution space.

Because of the constraint from the Kerr-Schild form, the solutions in NU gauge satisfying the real condition must be spherically symmetric. Then Birkhoff's theorem implies that the only vacuum solution is Schwarzschild. Nevertheless, we continue to derive a solution space by removing the real condition. There are many reasons to explore complex solutions of Einstein equation, in particular in the context of classical double copy. The complex solutions in the present work are real solutions in the split metric signature $(+,+,-,-)$. The self-dual sector in light-cone coordinates with split signature has very important applications in classical double copy \cite{Monteiro:2014cda,Berman:2018hwd,Monteiro:2020plf,Campiglia:2021srh}. Other applications of complex solutions can be found, e.g., in \cite{Ilderton:2017xbj,Adamo:2020syc,Witten:2021nzp}. In the stereographic coordinates, we find a self-consistent solution space. The news function and the initial data are constrained from the Kerr-Schild form. It is difficult to give a generic solution to the Kerr-Schild constraint. Alternatively, we provide two simple solutions for the news function and leading initial data. Those two cases correspond to two classes of solutions. It is worthwhile to emphasize that any solution of the news function and the initial data to the Kerr-Schild constraint will give a solution of vacuum Einstein equation. We also find an interesting solution in closed form. It is pp-wave solution and is a linear superposition of self-dual and anti-self-dual solutions. In the complex case, the news function is chiral. Consequently, it can not lead to a mass loss formula which is not surprising in the sense that the radiation taking away the energy from the system should satisfy real condition.

The organization of this paper is as follows: In the next section, we give a brief review of the NU gauge. In section 3, we combine the NU gauge with the Kerr-Schild form to fix the gauge conditions for the present work. Section 4 is devoted to the deviation of the solution space satisfying the gauge conditions. Exact solutions in the anti-self-dual or the self-dual sectors are also provided in this section. We close in the last section with some remarks.

\noindent
\textbf{Note added}: When writing up this paper, we became aware of \cite{Godazgar:2021iae,Adamo:2021dfg} with overlapping motivations with ours.

\section{The Newman-Unti gauge}
\label{NU}
The retarded coordinates $(u,r,z,\bz)$ is the most convenient choice of the coordinates system to work with physical fields at future null infinity where $(z,\bz)$ is the complex stereographic coordinates that is related to the usual angular variables $(\theta,\phi)$ by $z=\cot\frac\theta2 e^{i\phi}$. The NU gauge was originally introduced in the Newman-Penrose formalism \cite{Newman:1961qr} to study the asymptotic behavior of Weyl tensor and metric tensor. Adapted to metric formalism, the metric and inverse metric have the form \cite{Newman:1962cia}
\begin{equation}
g_{\mu\nu}=\begin{pmatrix}
g_{uu} & -1 & g_{uz} & g_{u\bz}\\
-1 & 0 & 0 & 0\\
 g_{uz} & 0 & g_{zz} & g_{z\bz}\\
 g_{u\bz} & 0 & g_{z\bz} & g_{\bz\bz}
\end{pmatrix}\ ,
\quad
g^{\mu\nu}=\begin{pmatrix}
0 & -1 & 0 & 0\\
-1 & g^{rr} & g^{rz} & g^{r\bz}\\
 0 & g^{rz} & g^{zz} & g^{z\bz}\\
 0 & g^{r\bz} & g^{z\bz} & g^{\bz\bz}
\end{pmatrix}\ .
\end{equation}
For computational simplicity, we map the celestial sphere at null infinity to a 2d plane. The connection between those two choices can be found, for instance, in \cite{Barnich:2016lyg,Compere:2016jwb,Ball:2018prg,Barnich:2021dta}. The line element of the Minkowski spacetime now is
\be\label{flat}
\td s^2=-2 \td u \td r + 2r^2 \td z \td \bz .
\ee
The asymptotic behaviors of the non-vanishing metric components preserving asymptotic flatness are required as
\be\begin{split}
&g_{uu}=\cO{(r^{-1})}\,,\quad g_{uz}=\cO(1)\,,\quad g_{u\bz}=\cO(1)\,,\\
&g_{zz}=\cO (r)\,,\quad g_{\bz\bz}=\cO (r)\,,\quad g_{z\bz}=r^2 + \cO(1)\,.
\end{split}\ee
To solve the Einstein equation, it is very useful to arrange all components as follows:
\begin{itemize}
\item
Four hypersurface equations:
\be
G_{r\mu}=0 \ .
\ee
\item
Two standard equations:
\be
G_{zz}=G_{\bz\bz}=0 \ .
\ee
\item
One trivial equation:
\be G_{z\bz}=0 \ .\ee
\item
Three supplementary equations:
\be
G_{uz}=G_{u\bz}=G_{uu}=0 \ .
\ee
\end{itemize}
The Bianchi identities can be written as
\be
\n_\mu {G^\mu}_\nu = \frac{1}{\sqrt{-g}}\p_\mu \left(\sqrt{-g} {G^\mu}_\nu \right) + \frac12 G_{\mu\rho} \p_\nu g^{\mu\rho}=0 \ .
\ee
When the hypersurface equations and standard equations are satisfied, the Bianchi identity $\n_\mu {G^\mu}_r=0$ yields
\be
G_{z\bz} \p_r g^{z\bz}=0 \ .
\ee
Hence $G_{z\bz}=0$ because of the boundary condition for $g_{z\bz}$, which means that the trivial equation is satisfied automatically. Then one obtains that $R=0$. The Bianchi identity $\n_\mu {G^\mu}_z=0$ reduces to $\p_r \left(\sqrt{-g} {G}_{uz} \right)=0$, noticing that the hypersurface equations and standard equations are satisfied. Similarly, the Bianchi identity $\n_\mu {G^\mu}_{\bz}=0$ yields $\p_r \left(\sqrt{-g} {G}_{u\bz} \right)=0$. The last Bianchi identity $\n_\mu {G^\mu}_u=0$ leads to $\p_r \left(\sqrt{-g} {G}_{uu} \right)=0$ when the rest equations are satisfied. Hence three supplementary equations are only left with only one order to solve in the asymptotic expansion.

\section{The Kerr-Schild form}

A metric of Kerr-Schild form is obtained from a deformation of a fixed background metric $\bar{g}_{\mu\nu}$. In the retarded coordinates $(u,r,z,\bz)$, we choose the background metric as the Minkowski spacetime \eqref{flat}. The deformation is made by a null vector field $k_\mu$ and a scalar function $\phi$ in the form
\be\label{KS}
g_{\mu\nu}=\eta_{\mu\nu} + \phi k_\mu k_\nu \ .
\ee
One can show that $k^\mu$ is null when contracted with both the full metric $g_{\mu\nu}$ and the Minkowski one $\eta_{\mu\nu}$. Therefore, the inverse metric is just
\be\label{KSinverse}
g^{\mu\nu}=\eta^{\mu\nu} - \phi k^\mu k^\nu \ .
\ee
Hence we can raise the index of $k$ using both $g^{\mu\nu}$ and $\eta^{\mu\nu}$. In the NU gauge, it is required that
\be
g_{rr}=g_{rz}=g_{r\bz}=0 \ , \quad g_{ur}=-1 \ .
\ee
Combining these conditions with the Kerr-Schild ansatz \eqref{KS}, one should require that $k_r=0$. Because $k_\mu$ is a null vector respect to $\eta^{\mu\nu}$, the solution of the remaining components is either $k_z=0$ or $k_{\bz}=0$. It is important to point out that such solution will not lead to a real metric. In other works, the only solution in the usual real angular variables $(\theta,\phi)$ is the trivial one $k_\theta=k_\phi=0$. However, it is still meaningful to continue the computation in the complex coordinates with the chiral solution. On the one hand, the self-dual metric fields in the retarded coordinates $(u,r,z,\bz)$ are exactly in this form. The self-dual gravity is a very important realization of the classical double copy \cite{Monteiro:2014cda,Berman:2018hwd,Monteiro:2020plf,Campiglia:2021srh}. Other important applications of self-dual gravity can be found, e.g., in \cite{Bianchi:1994gi,Duff:1994an,Ball:2021tmb} and references therein. On the other hand, this situation is somehow similar to the gluon three point amplitude at tree level. As the smallest or starting point amplitude in the BCFW recursion relation \cite{Britto:2005fq}, it is vanishing for real null momenta. Alternatively, one works with complex momenta where the helicity spinor $\lambda_i$ and $\tilde{\lambda}_i$ are independent and the three-gluon amplitudes are derived from either left-handed spinors or right-handed spionrs, see, e.g., \cite{Elvang:2013cua,Henn:2014yza} for a textbook treatment. We choose $k_{\bz}=0$ in the present work. Substituting this solution back to the metric, one obtains
\be\label{KSc1}
g_{u\bz}=0=g_{\bz\bz} \ .
\ee
The remaining components of the metric have the form
\be\label{metricform}
g_{uu}=\phi k_u k_u \ ,\quad g_{zz} = \phi k_z k_z \ , \quad g_{uz} = \phi k_u k_z \ .
\ee
Thus they should satisfy the relation as follows
\be\label{KSc2}
g_{uu}g_{zz}=g^2_{uz} \ .
\ee
The aim of the present work is to check the constraints from the Kerr-Schild form \eqref{KSc1} and \eqref{KSc2} on the solution space in NU gauge. Before doing so, we would like to point out that a double Kerr-Schild form
\be
g_{\mu\nu}=\eta_{\mu\nu} + \phi k_\mu k_\nu + \Psi l_\mu l_\nu \ ,
\ee
where $k^2=l^2=k\cdot l=0$, can not bring the other half copy of the metric. The NU gauge condition will fix both $k_r=0$ and $l_r=0$. The null conditions of $k_\mu$ and $l_\mu$ bring the same constraint as single Kerr-Schild form. Let us choose $k_{\bz}=0$ and $l_z=0$ to keep both components $g_{zz}$ and $g_{\bz\bz}$ in the metric. But the last condition that $k\cdot l=0$ eventually leads to $k_z l_{\bz}=0$. Hence we are back to the single Kerr-Schild form.

\section{The solution space}

Once the gauge condition from the Kerr-Schild is clarified, the solution of Einstein equation in the retarded coordinates can be derived in a straightforward way following the computation in \cite{Bondi:1962px,Sachs:1962wk,Barnich:2010eb,Conde:2016rom,Lu:2019jus}. We start with the hypersurface equations which determine the radial dependence of the metric from the initial data. The standard equations control the time evolutions of the initial data. Then the time evolution of the integration constants from the hypersurface equations are fixed by the supplementary equations. In this work, there is one more constraint from the Kerr-Schild form.

\subsection{Hypersurface equations}\label{hypersurface}

The hypersurface equations $G_{rr}=0$ and $G_{r\bz}=0$ are satisfied once the Kerr-Schild form is applied by imposing \eqref{KSc1}. The next one $G_{rz}=0$ yields
\be
\p_r \left(r^4 \p_r (\frac{g_{uz}}{r^2})\right)=r^2\p_r (\frac{\p_{\bz} g_{zz}}{r^2}) \ .
\ee
Hence $g_{uz}$ can be solved out from $g_{zz}$ as
\be\label{guz}
g_{uz} = \frac{N(u,z,\bz)}{r} + r^2 \int^{\infty}_r \td r' \frac{1}{{r'}^4} \int^{\infty}_{r'} \td r'' {r''}^2 \p_{r''} (\frac{\p_{\bz} g_{zz}}{{r''}^2}) \ ,
\ee
where function $N(u,z,\bz)$ is an integration constant in $r$. We continue with $G_{ur}=0$, from which one obtains
\be
\p_r (r g_{uu}) = \frac{1}{2r^2} \left[\p_r (r^2 \p_{\bz} g_{uz}) - \p_{\bz}^2 g_{zz} \right] \ .
\ee
Thus $g_{uu}$ is fixed as
\be\label{guu}
g_{uu}=\frac{M(u,z,\bz)}{r} - \frac1r \int^{\infty}_r  \td r' \frac{1}{2{r'}^2} \left[\p_{r'} ({r'}^2 \p_{\bz} g_{uz}) - \p_{\bz}^2 g_{zz} \right] \ ,
\ee
where $M(u,z,\bz)$ is another integration constant.

\subsection{Standard equations}

Equation $G_{\bz\bz}=0$ is satisfied automatically because of the Kerr-Schild form. The other standard equation leads to
\begin{multline}\label{u}
\p_u \p_r (\frac{g_{zz}}{r})=\frac{1}{2r^4} \bigg[ 4r g^2_{uz} + 2 g_{zz} \p_{\bz} g_{uz} + 4r^2 g_{zz} \p_r g_{uu} + r \p_{\bz} g_{zz} \p_r g_{uz} \\
 + r^3 (\p_r g_{uz})^2 - r \p_{\bz} g_{uz} \p_r g_{zz} - r^3 \p_r g_{uu} \p_r g_{zz}\\
  - 2r g_{zz} \p_r \p_{\bz} g_{uz} - g_{uz} (4 \p_{\bz} g_{zz} + 4r^2 \p_r g_{uz} - 2r \p_r \p_{\bz} g_{zz}) \\
+ 2r^3 \p_r \p_z g_{uz} + r^3 g_{zz} \p_r^2 g_{uu} - r g_{uu} (2 g_{zz} - 2 r \p_r g_{zz} + r^2 \p_r^2 g_{zz}) \bigg] \ .
\end{multline}
Clearly, there is no constraint at the order $\cO(r)$ of $g_{zz}$ from the standard equation. The field at this order is related to the news function in the system which indicates propagating degree of freedom.

\subsection{Solution space in series expansion}

Suppose that $g_{zz}$ is given as initial data as\footnote{The absent of $\bz$-dependence in $f_0(u,z)$ is to avoid logarithmic terms which will involve a polyhomogeneous solution space.}
\be\label{gzz}
g_{zz}=c(u,z,\bz) r + f_0(u,z) + \sum\limits^\infty_{i=1} \frac{f_i(u,z,\bz)}{r^i} \ .
\ee
Inserting the initial data into \eqref{guz} and \eqref{guu}, one obtains
\be
g_{uz}=\frac12 \p_z c + \frac{N}{r} - \sum\limits^\infty_{i=1} \frac{i+2}{i(i+3)} \frac{\p_{\bz} f_i } { r^{i+1}} \ ,
\ee
and
\be
g_{uu}=\frac{M}{r} - \frac{\p_{\bz} N}{2r^2} + \sum\limits^\infty_{i=1} \frac{1}{i(i+3)} \frac{\p_{\bz}^2 f_i }{r^{i+2}} \ .
\ee

\subsection{Supplementary equations}

With the series expansion, one can continue to solve the supplementary equations. From $G_{uz}=0$, we obtain
\be\label{Guz}
\p_u N = \frac13 \p_z M - \frac16 \p_z\p_{\bz}^2 c \ .
\ee
Next, $G_{u\bz}=0$ yields
\be\label{Guzb}
2\p_{\bz} M + \p_{\bz}^3 c = 0 \ .
\ee
The last supplementary equation $G_{uu}=0$ leads to
\be\label{Guu}
\p_u M =\frac12 \p_u \p_{\bz}^2 c \ .
\ee
It is of convenience to define
\be
M_0 = M -\frac12 \p_{\bz}^2 c \ .
\ee
Then \eqref{Guu} and \eqref{Guz} reduce to
\be
\p_u M_0 = 0 \ , \quad \p_u N = \frac13 \p_z M_0 \ .
\ee
Hence $M_0$ is independence of $u$ and $N= N_0(z,\bz) + \frac13 u \p_z M_0 $. So we do not have a mass loss formula in this system as $M_0$ is not changing. Moreover, \eqref{Guzb} becomes
\be
\p_{\bz} M_0 + \p_{\bz}^3 c =0 \ .
\ee
Since $M_0$ is independence of $u$, one can have that $\p_u\p_{\bz}^3 c = 0$. Therefore, a generic solution to $c$ should have the form
\be
c={\tilde c}(z,\bz) + c_0(u,z) + c_1(u,z) \bz + c_2(u,z) \bz^2 \ .
\ee
Hence
\be
M_0=\xbar{M}(z) - \p_{\bz}^2\tilde{c} \ ,\quad M = \xbar{M} - \frac12\p_{\bz}^2\tilde{c} + c_2 \ , \quad N=N_0 + \frac13 u (\p_z \xbar M - \p_z \p_{\bz}^2\tilde{c}) \ .
\ee

\subsection{The Kerr-Shcild constraint}

The first two constraints in \eqref{KSc1} from the Kerr-Schild form have been implemented when solving the Einstein equation. The last one in \eqref{KSc2} can be considered as a constraint on the initial data. In the series expansion \eqref{gzz}, the fields from each order are constrained from both the standard equation \eqref{u} and the Kerr-Schild form \eqref{KSc2}. For the first order, the news aspect $c$ is only constrained from the Kerr-Schild form as
\be
c M =\frac14 (\p_{\bz} c)^2 \ ,
\ee
For the second order, one obtains
\begin{align}
&\p_u f_0=0 \ , \\
&f_0 M = N \p_{\bz} c + \frac12 c \p_{\bz} N \ .
\end{align}
Clearly, the first equation shows that $f_0$ has only $z$ dependence. Hence the second equation will lead to another constraint on $c$. Though it is not easy to give a generic solution for $c$, one can easily find some non trivial solutions, e.g.,
\be\label{special}
f_0=M_0=N_0=\tilde{c}=c_0=c_1=0 \ , \quad M=c_2(u,z) \ , \quad c=c_2(u,z)\bz^2 \ .
\ee
and
\be
M=\tilde{c}=c_1=c_2=0 \ , \quad N=N_0(z) \ , \quad c=c_0(u,z) \ .
\ee
We can continue to trace higher orders. For every $f_i(i\geq1)$, there are two differential equations. Considering $f_1$ as example, one has
\begin{align}
&2 f_0 \p_{\bz} N - 3 \p_{\bz} f_1 \p_{\bz} c - c \p_{\bz}^2 f_1 - 4 f_1 M + 4 N^2 = 0 \ ,\\
&\p_u f_1 = \frac18 c(2M-\p_{\bz}^2 c) + \frac12 \p_z N \ .
\end{align}
Again, a generic solution for $f_1$ is very to find once the other fields in the equations are known. One can show that there is non trivial solution when the other fields are given as \eqref{special}. A simple one is $f_1=\frac{\tilde{f}_1(z)}{\bz}$ where $\tilde{f}_1(z)$ is an arbitrary function of $z$. Hence, one can expect that there should not be more constraint on the news aspect $c$ from the equations at higher orders. Since the dynamics is only encoded in the news aspect $c$, the equations from higher orders will not bring any new information.

\subsection{Exact solution}

There is a simple but very interesting solution in closed form in the solution space. The metric is given by
\be\label{exact}
\td s^2=-2\td u\td r + 2 r^2 \td z \td \bz + r [c(u,z) + f(r,z)]\td z^2 \ .
\ee
This solution is non-trivial in the sense that it has non vanishing components of Riemann tensor which are
\be
R_{uzuz}=-\frac12 r \p_u^2 c \ , \quad R_{rzrz}=-\frac12 r \p_r^2 f \ .
\ee
The square of the Riemann tensor $R_{\mu\nu\alpha\beta}R^{\mu\nu\alpha\beta}$ is zero. The vector $k=\frac{\p}{\p \bz}$ is null and tangent to
an expansion-free, shear-free and twist-free null geodesic congruence, i.e., satisfying $\nabla_\mu k_\nu = 0$. So this solution belongs to the class of pp-wave solutions following the definition in \cite{Ehlers:1962zz,Griffiths:2009dfa}. Interestingly, this pp-wave solution is asymptotically flat when the asymptotic behavior of $f(r,z)$ is $\cO(\frac1r)$ which is a direct consequence of the NU setup in section \ref{NU} with a special boundary choice \eqref{flat}, i.e., a 2d plane. Coordinate $z$ labels the null wave surfaces and coordinates $u$ and $r$ span the wave surfaces. There is another null direction $l=\frac{\p}{\p r}$ that has expansion and shear. With the following change of coordinates
\be
\bz=\rho \ , \quad z=-v \ , \quad r=\frac1\zeta \ ,\quad u=\bar\zeta \ ,
\ee
the line element \eqref{exact} becomes
\be
\td s^2=\frac{1}{\zeta^2} \left[-2\td v \td \rho - 2H(v,\zeta,\bar\zeta)\td v^2 + 2 \td \zeta \td \bar\zeta \right],
\ee
where $H(v,\zeta,\bar\zeta)=-\frac12 \zeta [c(v,\bar\zeta) + f(v,\zeta)]$. This metric is conformal to the Brinkmann form \cite{Griffiths:2009dfa} with a complex conformal factor. This solution might be considered as a generalization of the Brinkmann form to the complex case. Writing the line element \eqref{exact} in a more generic form as
\be\label{general}
\td s^2=-2\td u\td r + 2 r^2 \td z \td \bz + r H(u,r,z)\td z^2 \ ,
\ee
the vacuum Einstein yields
\be\label{mainequation}
\p_u\p_r H(u,r,z)=0 \ .
\ee
The precise choice in \eqref{exact} is just the general solution to this equation.

By choosing four null basis as
\be
\label{NP}
e^1_\mu=(0,1,0,0) \ , e^2_\mu=(1,0,0,0) \ , e^3_\mu=(0,0,r,0) \ , e^4_\mu=(0,0,\frac{H}{2},r) \ ,
\ee
the solution can be described in a Newman-Penrose formalism. The non-vanishing Newman-Penrose variables are
\be\begin{split}
&\Psi_0=\frac{\p_r^2 H}{2r} \ ,\quad  \xbar\Psi_4=\frac{\p_u^2 H}{2r} \ ,\\
&\rho=\xbar\rho=\frac1r \ , \quad \sigma=-\frac{H-r\p_r H}{2r^2} \ , \quad \xbar\lambda=-\frac{\p_u H}{2r} \ .\end{split}
\ee

Since \eqref{mainequation} is a linear equation, a new solution can be obtained simply by superposing distinct solutions with different expressions for $H$ just like the case of pp-wave in Brinkmann form. An interesting feature of the solution \eqref{exact} is the fact that it can be considered as a linear superposition of self-dual and anti-self-dual solutions. The case $c=0$ and $f\neq0$ represents a self-dual solution while the other case $c\neq0$ and $f=0$ represents an anti-self-dual solution. To see that, we define the spinorial tetrad as
\be
{\sigma_\mu}_{A\dot{A}}=\sigma^a_{A\dot{A}} e^b_\mu \eta_{ab} \ ,
\ee
where
\be
\eta_{ab}=\eta^{ab}=\begin{pmatrix}
0 & -1 & 0 & 0  \\ -1 & 0 & 0 & 0  \\ 0 & 0 & 0 & 1  \\ 0 & 0 & 1 & 0
\end{pmatrix}\ , \quad \sigma^a=\frac12(\sigma^2 + i \sigma^3, \sigma^2 - i \sigma^3,\mathbb{1}-\sigma^1,\mathbb{1}+\sigma^1) \ ,
\ee
$\sigma^k(k = 1,2,3)$ are the Pauli matrices and the tetrad is chosen as \eqref{NP}. In our convention, $\varepsilon^{12}=1$. One can easily verify that
\be
{\sigma_\mu}_{A\dot{A}}{\sigma_\nu}_{B\dot{B}}\varepsilon^{AB}\varepsilon^{\dot{A}\dot{B}}=g_{\mu\nu} \ .
\ee
The spinoral form of the Weyl tensor is
\be
W_{A\dot{A}B\dot{B}C\dot{C}D\dot{D}}=C_{ABCD} \varepsilon_{\dot{A}\dot{B}}\varepsilon_{\dot{C}\dot{D}} + \xbar C_{\dot{A}\dot{B}\dot{C}\dot{D}}\varepsilon_{AB}\varepsilon_{CD} \ .
\ee
$C_{ABCD}$ and $\xbar C_{\dot{A}\dot{B}\dot{C}\dot{D}}$ are totally symmetric and they represent the anti-self-dual and self-dual parts of the curvature respectively. For a real metric, they are complex conjugation of each other. Direct computation regarding to the solution \eqref{exact} gives
\be
C_{1111}=C_{1122}=C_{2222}=\frac{1}{8r} \p_u^2 c\ , \quad C_{1112}=C_{1222}=-\frac{1}{8r} \p_u^2 c\ ,
\ee
and
\be
\xbar C_{1111}=\xbar C_{1122}=\xbar C_{2222}=\frac{1}{8r} \p_r^2 f \ , \quad \xbar C_{1112}=\xbar C_{1222}=-\frac{1}{8r} \p_r^2 f \ .
\ee

\section{Discussion}

We obtain a generic solution space in series expansion near null infinity in the NU gauge satisfying the Kerr-Schild form. The solutions only exist in complex coordinates which should correspond to a complex classical double copy. Interesting exact solutions in the anti-self-dual and the self-dual sectors can be read out from the solution space. As the single copy or zeroth copy version of the solution space, the null vector $k_\mu$ and the scalar field $\phi$ in the Kerr-Schild form can be easily derived from \eqref{metricform}.

The solution space is self-contained. But the result itself somehow indicates the incompatibility between the NU gauge and Kerr-Schild form. The only non trivial real solution in NU gauge satisfying the Kerr-Schild form is the spherically symmetric one. With such shortcoming, applications of the present result in the triangle relation are now less clear. For instance, what is a memory effect in the Kerr-Schild form and what is the interpretation of the memory effect from the point of view of classical double copy structure. We do not yet have a clear answer to those questions and leave them for future investigation.

\section*{Acknowledgments}

The authors thank Zhengwen Liu, Hong L\"{u}, Jun-Bao Wu, and Xiaoning Wu for useful discussions. This work is supported in part by the National Natural Science Foundation of China under Grant No. 11905156 and No. 11935009.

\providecommand{\href}[2]{#2}\begingroup\raggedright\endgroup


\providecommand{\href}[2]{#2}\begingroup\raggedright\begin{thebibliography}{}

\end{thebibliography}\endgroup


\begin{thebibliography}{10}

\bibitem{Bern:2008qj}
Z.~Bern, J.~J.~M. Carrasco, and H.~Johansson, ``{New Relations for Gauge-Theory
  Amplitudes},'' \href{http://dx.doi.org/10.1103/PhysRevD.78.085011}{{\em Phys.
  Rev. D} {\bfseries 78} (2008) 085011},
  \href{http://arxiv.org/abs/0805.3993}{{\ttfamily arXiv:0805.3993 [hep-ph]}}.

\bibitem{Bern:2010ue}
Z.~Bern, J.~J.~M. Carrasco, and H.~Johansson, ``{Perturbative Quantum Gravity
  as a Double Copy of Gauge Theory},''
  \href{http://dx.doi.org/10.1103/PhysRevLett.105.061602}{{\em Phys. Rev.
  Lett.} {\bfseries 105} (2010) 061602},
  \href{http://arxiv.org/abs/1004.0476}{{\ttfamily arXiv:1004.0476 [hep-th]}}.

\bibitem{Bern:2019prr}
Z.~Bern, J.~J. Carrasco, M.~Chiodaroli, H.~Johansson, and R.~Roiban, ``{The
  Duality Between Color and Kinematics and its Applications},''
  \href{http://arxiv.org/abs/1909.01358}{{\ttfamily arXiv:1909.01358
  [hep-th]}}.

\bibitem{Monteiro:2014cda}
R.~Monteiro, D.~O'Connell, and C.~D. White, ``{Black holes and the double
  copy},'' \href{http://dx.doi.org/10.1007/JHEP12(2014)056}{{\em JHEP}
  {\bfseries 12} (2014) 056}, \href{http://arxiv.org/abs/1410.0239}{{\ttfamily
  arXiv:1410.0239 [hep-th]}}.

\bibitem{Luna:2015paa}
A.~Luna, R.~Monteiro, D.~O'Connell, and C.~D. White, ``{The classical double
  copy for Taub\textendash{}NUT spacetime},''
  \href{http://dx.doi.org/10.1016/j.physletb.2015.09.021}{{\em Phys. Lett. B}
  {\bfseries 750} (2015) 272--277},
  \href{http://arxiv.org/abs/1507.01869}{{\ttfamily arXiv:1507.01869
  [hep-th]}}.

\bibitem{Luna:2016due}
A.~Luna, R.~Monteiro, I.~Nicholson, D.~O'Connell, and C.~D. White, ``{The
  double copy: Bremsstrahlung and accelerating black holes},''
  \href{http://dx.doi.org/10.1007/JHEP06(2016)023}{{\em JHEP} {\bfseries 06}
  (2016) 023}, \href{http://arxiv.org/abs/1603.05737}{{\ttfamily
  arXiv:1603.05737 [hep-th]}}.

\bibitem{White:2016jzc}
C.~D. White, ``{Exact solutions for the biadjoint scalar field},''
  \href{http://dx.doi.org/10.1016/j.physletb.2016.10.052}{{\em Phys. Lett. B}
  {\bfseries 763} (2016) 365--369},
  \href{http://arxiv.org/abs/1606.04724}{{\ttfamily arXiv:1606.04724
  [hep-th]}}.

\bibitem{Bahjat-Abbas:2017htu}
N.~Bahjat-Abbas, A.~Luna, and C.~D. White, ``{The Kerr-Schild double copy in
  curved spacetime},'' \href{http://dx.doi.org/10.1007/JHEP12(2017)004}{{\em
  JHEP} {\bfseries 12} (2017) 004},
  \href{http://arxiv.org/abs/1710.01953}{{\ttfamily arXiv:1710.01953
  [hep-th]}}.

\bibitem{Carrillo-Gonzalez:2017iyj}
M.~Carrillo-Gonz\'alez, R.~Penco, and M.~Trodden, ``{The classical double copy
  in maximally symmetric spacetimes},''
  \href{http://dx.doi.org/10.1007/JHEP04(2018)028}{{\em JHEP} {\bfseries 04}
  (2018) 028}, \href{http://arxiv.org/abs/1711.01296}{{\ttfamily
  arXiv:1711.01296 [hep-th]}}.

\bibitem{Goldberger:2017ogt}
W.~D. Goldberger, J.~Li, and S.~G. Prabhu, ``{Spinning particles, axion
  radiation, and the classical double copy},''
  \href{http://dx.doi.org/10.1103/PhysRevD.97.105018}{{\em Phys. Rev. D}
  {\bfseries 97} no.~10, (2018) 105018},
  \href{http://arxiv.org/abs/1712.09250}{{\ttfamily arXiv:1712.09250
  [hep-th]}}.

\bibitem{Li:2018qap}
J.~Li and S.~G. Prabhu, ``{Gravitational radiation from the classical spinning
  double copy},'' \href{http://dx.doi.org/10.1103/PhysRevD.97.105019}{{\em
  Phys. Rev. D} {\bfseries 97} no.~10, (2018) 105019},
  \href{http://arxiv.org/abs/1803.02405}{{\ttfamily arXiv:1803.02405
  [hep-th]}}.

\bibitem{Lee:2018gxc}
K.~Lee, ``{Kerr-Schild Double Field Theory and Classical Double Copy},''
  \href{http://dx.doi.org/10.1007/JHEP10(2018)027}{{\em JHEP} {\bfseries 10}
  (2018) 027}, \href{http://arxiv.org/abs/1807.08443}{{\ttfamily
  arXiv:1807.08443 [hep-th]}}.

\bibitem{Berman:2018hwd}
D.~S. Berman, E.~Chac\'on, A.~Luna, and C.~D. White, ``{The self-dual classical
  double copy, and the Eguchi-Hanson instanton},''
  \href{http://dx.doi.org/10.1007/JHEP01(2019)107}{{\em JHEP} {\bfseries 01}
  (2019) 107}, \href{http://arxiv.org/abs/1809.04063}{{\ttfamily
  arXiv:1809.04063 [hep-th]}}.

\bibitem{Gurses:2018ckx}
M.~Gurses and B.~Tekin, ``{Classical Double Copy: Kerr-Schild-Kundt metrics
  from Yang-Mills Theory},''
  \href{http://dx.doi.org/10.1103/PhysRevD.98.126017}{{\em Phys. Rev. D}
  {\bfseries 98} no.~12, (2018) 126017},
  \href{http://arxiv.org/abs/1810.03411}{{\ttfamily arXiv:1810.03411 [gr-qc]}}.

\bibitem{Luna:2018dpt}
A.~Luna, R.~Monteiro, I.~Nicholson, and D.~O'Connell, ``{Type D Spacetimes and
  the Weyl Double Copy},''
  \href{http://dx.doi.org/10.1088/1361-6382/ab03e6}{{\em Class. Quant. Grav.}
  {\bfseries 36} (2019) 065003},
  \href{http://arxiv.org/abs/1810.08183}{{\ttfamily arXiv:1810.08183
  [hep-th]}}.

\bibitem{CarrilloGonzalez:2019gof}
M.~Carrillo~Gonz\'alez, B.~Melcher, K.~Ratliff, S.~Watson, and C.~D. White,
  ``{The classical double copy in three spacetime dimensions},''
  \href{http://dx.doi.org/10.1007/JHEP07(2019)167}{{\em JHEP} {\bfseries 07}
  (2019) 167}, \href{http://arxiv.org/abs/1904.11001}{{\ttfamily
  arXiv:1904.11001 [hep-th]}}.

\bibitem{Cho:2019ype}
W.~Cho and K.~Lee, ``{Heterotic Kerr-Schild Double Field Theory and Classical
  Double Copy},'' \href{http://dx.doi.org/10.1007/JHEP07(2019)030}{{\em JHEP}
  {\bfseries 07} (2019) 030}, \href{http://arxiv.org/abs/1904.11650}{{\ttfamily
  arXiv:1904.11650 [hep-th]}}.

\bibitem{PV:2019uuv}
A.~P.~V. and A.~Manu, ``{Classical double copy from Color Kinematics duality: A
  proof in the soft limit},''
  \href{http://dx.doi.org/10.1103/PhysRevD.101.046014}{{\em Phys. Rev. D}
  {\bfseries 101} no.~4, (2020) 046014},
  \href{http://arxiv.org/abs/1907.10021}{{\ttfamily arXiv:1907.10021
  [hep-th]}}.

\bibitem{Bautista:2019evw}
Y.~F. Bautista and A.~Guevara, ``{On the Double Copy for Spinning Matter},''
  \href{http://arxiv.org/abs/1908.11349}{{\ttfamily arXiv:1908.11349
  [hep-th]}}.

\bibitem{Kim:2019jwm}
K.~Kim, K.~Lee, R.~Monteiro, I.~Nicholson, and D.~Peinador~Veiga, ``{The
  Classical Double Copy of a Point Charge},''
  \href{http://dx.doi.org/10.1007/JHEP02(2020)046}{{\em JHEP} {\bfseries 02}
  (2020) 046}, \href{http://arxiv.org/abs/1912.02177}{{\ttfamily
  arXiv:1912.02177 [hep-th]}}.

\bibitem{Bah:2019sda}
I.~Bah, R.~Dempsey, and P.~Weck, ``{Kerr-Schild Double Copy and Complex
  Worldlines},'' \href{http://dx.doi.org/10.1007/JHEP02(2020)180}{{\em JHEP}
  {\bfseries 02} (2020) 180}, \href{http://arxiv.org/abs/1910.04197}{{\ttfamily
  arXiv:1910.04197 [hep-th]}}.

\bibitem{Arkani-Hamed:2019ymq}
N.~Arkani-Hamed, Y.-t. Huang, and D.~O'Connell, ``{Kerr black holes as
  elementary particles},''
  \href{http://dx.doi.org/10.1007/JHEP01(2020)046}{{\em JHEP} {\bfseries 01}
  (2020) 046}, \href{http://arxiv.org/abs/1906.10100}{{\ttfamily
  arXiv:1906.10100 [hep-th]}}.

\bibitem{Lescano:2020nve}
E.~Lescano and J.~A. Rodr\'\i{}guez, ``{$ \mathcal{N} $ = 1 supersymmetric
  Double Field Theory and the generalized Kerr-Schild ansatz},''
  \href{http://dx.doi.org/10.1007/JHEP10(2020)148}{{\em JHEP} {\bfseries 10}
  (2020) 148}, \href{http://arxiv.org/abs/2002.07751}{{\ttfamily
  arXiv:2002.07751 [hep-th]}}.

\bibitem{Gumus:2020hbb}
M.~K. Gumus and G.~Alkac, ``{More on the classical double copy in three
  spacetime dimensions},''
  \href{http://dx.doi.org/10.1103/PhysRevD.102.024074}{{\em Phys. Rev. D}
  {\bfseries 102} no.~2, (2020) 024074},
  \href{http://arxiv.org/abs/2006.00552}{{\ttfamily arXiv:2006.00552
  [hep-th]}}.

\bibitem{Elor:2020nqe}
G.~Elor, K.~Farnsworth, M.~L. Graesser, and G.~Herczeg, ``{The Newman-Penrose
  Map and the Classical Double Copy},''
  \href{http://dx.doi.org/10.1007/JHEP12(2020)121}{{\em JHEP} {\bfseries 12}
  (2020) 121}, \href{http://arxiv.org/abs/2006.08630}{{\ttfamily
  arXiv:2006.08630 [hep-th]}}.

\bibitem{Easson:2020esh}
D.~A. Easson, C.~Keeler, and T.~Manton, ``{Classical double copy of nonsingular
  black holes},'' \href{http://dx.doi.org/10.1103/PhysRevD.102.086015}{{\em
  Phys. Rev. D} {\bfseries 102} no.~8, (2020) 086015},
  \href{http://arxiv.org/abs/2007.16186}{{\ttfamily arXiv:2007.16186 [gr-qc]}}.

\bibitem{Godazgar:2020zbv}
H.~Godazgar, M.~Godazgar, R.~Monteiro, D.~P. Veiga, and C.~N. Pope, ``{Weyl
  Double Copy for Gravitational Waves},''
  \href{http://dx.doi.org/10.1103/PhysRevLett.126.101103}{{\em Phys. Rev.
  Lett.} {\bfseries 126} no.~10, (2021) 101103},
  \href{http://arxiv.org/abs/2010.02925}{{\ttfamily arXiv:2010.02925
  [hep-th]}}.

\bibitem{Berman:2020xvs}
D.~S. Berman, K.~Kim, and K.~Lee, ``{The classical double copy for M-theory
  from a Kerr-Schild ansatz for exceptional field theory},''
  \href{http://dx.doi.org/10.1007/JHEP04(2021)071}{{\em JHEP} {\bfseries 04}
  (2021) 071}, \href{http://arxiv.org/abs/2010.08255}{{\ttfamily
  arXiv:2010.08255 [hep-th]}}.

\bibitem{Prabhu:2020avf}
S.~G. Prabhu, ``{The classical double copy in curved spacetimes: Perturbative
  Yang-Mills from the bi-adjoint scalar},''
  \href{http://arxiv.org/abs/2011.06588}{{\ttfamily arXiv:2011.06588
  [hep-th]}}.

\bibitem{White:2020sfn}
C.~D. White, ``{Twistorial Foundation for the Classical Double Copy},''
  \href{http://dx.doi.org/10.1103/PhysRevLett.126.061602}{{\em Phys. Rev.
  Lett.} {\bfseries 126} no.~6, (2021) 061602},
  \href{http://arxiv.org/abs/2012.02479}{{\ttfamily arXiv:2012.02479
  [hep-th]}}.

\bibitem{Monteiro:2020plf}
R.~Monteiro, D.~O'Connell, D.~P. Veiga, and M.~Sergola, ``{Classical solutions
  and their double copy in split signature},''
  \href{http://dx.doi.org/10.1007/JHEP05(2021)268}{{\em JHEP} {\bfseries 05}
  (2021) 268}, \href{http://arxiv.org/abs/2012.11190}{{\ttfamily
  arXiv:2012.11190 [hep-th]}}.

\bibitem{Lescano:2021ooe}
E.~Lescano and J.~A. Rodr\'\i{}guez, ``{Higher-derivative heterotic Double
  Field Theory and classical double copy},''
  \href{http://dx.doi.org/10.1007/JHEP07(2021)072}{{\em JHEP} {\bfseries 07}
  (2021) 072}, \href{http://arxiv.org/abs/2101.03376}{{\ttfamily
  arXiv:2101.03376 [hep-th]}}.

\bibitem{Campiglia:2021srh}
M.~Campiglia and S.~Nagy, ``{A double copy for asymptotic symmetries in the
  self-dual sector},'' \href{http://dx.doi.org/10.1007/JHEP03(2021)262}{{\em
  JHEP} {\bfseries 03} (2021) 262},
  \href{http://arxiv.org/abs/2102.01680}{{\ttfamily arXiv:2102.01680
  [hep-th]}}.

\bibitem{Alkac:2021bav}
G.~Alkac, M.~K. Gumus, and M.~Tek, ``{The Kerr-Schild Double Copy in Lifshitz
  Spacetime},'' \href{http://dx.doi.org/10.1007/JHEP05(2021)214}{{\em JHEP}
  {\bfseries 05} (2021) 214}, \href{http://arxiv.org/abs/2103.06986}{{\ttfamily
  arXiv:2103.06986 [hep-th]}}.

\bibitem{Chacon:2021wbr}
E.~Chac\'on, S.~Nagy, and C.~D. White, ``{The Weyl double copy from twistor
  space},'' \href{http://dx.doi.org/10.1007/JHEP05(2021)239}{{\em JHEP}
  {\bfseries 05} (2021) 2239},
  \href{http://arxiv.org/abs/2103.16441}{{\ttfamily arXiv:2103.16441
  [hep-th]}}.

\bibitem{Alkac:2021seh}
G.~Alkac, M.~K. Gumus, and M.~A. Olpak, ``{Kerr-Schild double copy of the
  Coulomb solution in three dimensions},''
  \href{http://dx.doi.org/10.1103/PhysRevD.104.044034}{{\em Phys. Rev. D}
  {\bfseries 104} no.~4, (2021) 044034},
  \href{http://arxiv.org/abs/2105.11550}{{\ttfamily arXiv:2105.11550
  [hep-th]}}.

\bibitem{Angus:2021zhy}
S.~Angus, K.~Cho, and K.~Lee, ``{The classical double copy for half-maximal
  supergravities and T-duality},''
  \href{http://dx.doi.org/10.1007/JHEP10(2021)211}{{\em JHEP} {\bfseries 10}
  (2021) 211}, \href{http://arxiv.org/abs/2105.12857}{{\ttfamily
  arXiv:2105.12857 [hep-th]}}.

\bibitem{Chacon:2021hfe}
E.~Chac\'on, A.~Luna, and C.~D. White, ``{The double copy of the multipole
  expansion},'' \href{http://arxiv.org/abs/2108.07702}{{\ttfamily
  arXiv:2108.07702 [hep-th]}}.

\bibitem{Diaz-Jaramillo:2021wtl}
F.~Diaz-Jaramillo, O.~Hohm, and J.~Plefka, ``{Double Field Theory as the Double
  Copy of Yang-Mills},'' \href{http://arxiv.org/abs/2109.01153}{{\ttfamily
  arXiv:2109.01153 [hep-th]}}.

\bibitem{Gonzo:2021drq}
R.~Gonzo and C.~Shi, ``{Geodesics from classical double copy},''
  \href{http://dx.doi.org/10.1103/PhysRevD.104.105012}{{\em Phys. Rev. D}
  {\bfseries 104} no.~10, (2021) 105012},
  \href{http://arxiv.org/abs/2109.01072}{{\ttfamily arXiv:2109.01072
  [hep-th]}}.

\bibitem{Cho:2021nim}
K.~Cho, K.~Kim, and K.~Lee, ``{The Off-Shell Recursion for Gravity and the
  Classical Double Copy for currents},''
  \href{http://arxiv.org/abs/2109.06392}{{\ttfamily arXiv:2109.06392
  [hep-th]}}.

\bibitem{Godazgar:2021iae}
H.~Godazgar, M.~Godazgar, R.~Monteiro, D.~Peinador~Veiga, and C.~N. Pope,
  ``{Asymptotic Weyl double copy},''
  \href{http://dx.doi.org/10.1007/JHEP11(2021)126}{{\em JHEP} {\bfseries 11}
  (2021) 126}, \href{http://arxiv.org/abs/2109.07866}{{\ttfamily
  arXiv:2109.07866 [hep-th]}}.

\bibitem{Adamo:2021dfg}
T.~Adamo and U.~Kol, ``{Classical double copy at null infinity},''
  \href{http://arxiv.org/abs/2109.07832}{{\ttfamily arXiv:2109.07832
  [hep-th]}}.

\bibitem{Stephani:2003tm}
H.~Stephani, D.~Kramer, M.~A.~H. MacCallum, C.~Hoenselaers, and E.~Herlt,
  \href{http://dx.doi.org/10.1017/CBO9780511535185}{{\em {Exact solutions of
  Einstein's field equations}}}.
\newblock Cambridge Monographs on Mathematical Physics. Cambridge Univ. Press,
  Cambridge, 2003.

\bibitem{Bondi:1962px}
H.~Bondi, M.~G.~J. van~der Burg, and A.~W.~K. Metzner, ``{Gravitational waves
  in general relativity. 7. Waves from axisymmetric isolated systems},''
\href{http://dx.doi.org/10.1098/rspa.1962.0161}{{\em Proc. Roy. Soc. Lond.}
  {\bfseries A269} (1962) 21--52}.

\bibitem{Chong:2004na}
Z.~W. Chong, M.~Cvetic, H.~Lu, and C.~N. Pope, ``{Charged rotating black holes
  in four-dimensional gauged and ungauged supergravities},''
  \href{http://dx.doi.org/10.1016/j.nuclphysb.2005.03.034}{{\em Nucl. Phys. B}
  {\bfseries 717} (2005) 246--271},
  \href{http://arxiv.org/abs/hep-th/0411045}{{\ttfamily arXiv:hep-th/0411045}}.

\bibitem{Strominger:2017zoo}
A.~Strominger, ``{Lectures on the Infrared Structure of Gravity and Gauge
  Theory},''
\href{http://arxiv.org/abs/1703.05448}{{\ttfamily arXiv:1703.05448 [hep-th]}}.

\bibitem{Newman:1962cia}
E.~T. Newman and T.~W.~J. Unti, ``{Behavior of Asymptotically Flat Empty
  Spaces},''
\href{http://dx.doi.org/10.1063/1.1724303}{{\em J. Math. Phys.} {\bfseries 3}
  no.~5, (1962) 891}.

\bibitem{Ilderton:2017xbj}
A.~Ilderton and D.~Seipt, ``{Backreaction on background fields: A coherent
  state approach},'' \href{http://dx.doi.org/10.1103/PhysRevD.97.016007}{{\em
  Phys. Rev. D} {\bfseries 97} no.~1, (2018) 016007},
  \href{http://arxiv.org/abs/1709.10085}{{\ttfamily arXiv:1709.10085
  [hep-th]}}.

\bibitem{Adamo:2020syc}
T.~Adamo, L.~Mason, and A.~Sharma, ``{MHV scattering of gluons and gravitons in
  chiral strong fields},''
  \href{http://dx.doi.org/10.1103/PhysRevLett.125.041602}{{\em Phys. Rev.
  Lett.} {\bfseries 125} no.~4, (2020) 041602},
  \href{http://arxiv.org/abs/2003.13501}{{\ttfamily arXiv:2003.13501
  [hep-th]}}.

\bibitem{Witten:2021nzp}
E.~Witten, ``{A Note On Complex Spacetime Metrics},''
  \href{http://arxiv.org/abs/2111.06514}{{\ttfamily arXiv:2111.06514
  [hep-th]}}.

\bibitem{Newman:1961qr}
E.~Newman and R.~Penrose, ``{An Approach to gravitational radiation by a method
  of spin coefficients},''
\href{http://dx.doi.org/10.1063/1.1724257}{{\em J. Math. Phys.} {\bfseries 3}
  (1962) 566--578}.

\bibitem{Barnich:2016lyg}
G.~Barnich and C.~Troessaert, ``{Finite BMS transformations},''
  \href{http://dx.doi.org/10.1007/JHEP03(2016)167}{{\em JHEP} {\bfseries 03}
  (2016) 167}, \href{http://arxiv.org/abs/1601.04090}{{\ttfamily
  arXiv:1601.04090 [gr-qc]}}.

\bibitem{Compere:2016jwb}
G.~Comp\`{e}re and J.~Long, ``{Vacua of the gravitational field},''
  \href{http://dx.doi.org/10.1007/JHEP07(2016)137}{{\em JHEP} {\bfseries 07}
  (2016) 137}, \href{http://arxiv.org/abs/1601.04958}{{\ttfamily
  arXiv:1601.04958 [hep-th]}}.

\bibitem{Ball:2018prg}
A.~Ball, M.~Pate, A.-M. Raclariu, A.~Strominger, and R.~Venugopalan,
  ``{Measuring color memory in a color glass condensate at
  electron\textendash{}ion colliders},''
  \href{http://dx.doi.org/10.1016/j.aop.2019.04.010}{{\em Annals Phys.}
  {\bfseries 407} (2019) 15--28},
  \href{http://arxiv.org/abs/1805.12224}{{\ttfamily arXiv:1805.12224
  [hep-ph]}}.

\bibitem{Barnich:2021dta}
G.~Barnich and R.~Ruzziconi, ``{Coadjoint representation of the BMS group on
  celestial Riemann surfaces},''
  \href{http://arxiv.org/abs/2103.11253}{{\ttfamily arXiv:2103.11253 [gr-qc]}}.

\bibitem{Bianchi:1994gi}
M.~Bianchi, F.~Fucito, G.~C. Rossi, and M.~Martellini, ``{ALE instantons in
  string effective theory},''
  \href{http://dx.doi.org/10.1016/0550-3213(94)00552-P}{{\em Nucl. Phys. B}
  {\bfseries 440} (1995) 129--170},
  \href{http://arxiv.org/abs/hep-th/9409037}{{\ttfamily arXiv:hep-th/9409037}}.

\bibitem{Duff:1994an}
M.~J. Duff, R.~R. Khuri, and J.~X. Lu, ``{String solitons},''
  \href{http://dx.doi.org/10.1016/0370-1573(95)00002-X}{{\em Phys. Rept.}
  {\bfseries 259} (1995) 213--326},
  \href{http://arxiv.org/abs/hep-th/9412184}{{\ttfamily arXiv:hep-th/9412184}}.

\bibitem{Ball:2021tmb}
A.~Ball, S.~Narayanan, J.~Salzer, and A.~Strominger, ``{Perturbatively Exact
  $w_{1+\infty}$ Asymptotic Symmetry of Quantum Self-Dual Gravity},''
  \href{http://arxiv.org/abs/2111.10392}{{\ttfamily arXiv:2111.10392
  [hep-th]}}.

\bibitem{Britto:2005fq}
R.~Britto, F.~Cachazo, B.~Feng, and E.~Witten, ``{Direct proof of tree-level
  recursion relation in Yang-Mills theory},''
  \href{http://dx.doi.org/10.1103/PhysRevLett.94.181602}{{\em Phys. Rev. Lett.}
  {\bfseries 94} (2005) 181602},
  \href{http://arxiv.org/abs/hep-th/0501052}{{\ttfamily arXiv:hep-th/0501052}}.

\bibitem{Elvang:2013cua}
H.~Elvang and Y.-t. Huang, ``{Scattering Amplitudes},''
  \href{http://arxiv.org/abs/1308.1697}{{\ttfamily arXiv:1308.1697 [hep-th]}}.

\bibitem{Henn:2014yza}
J.~M. Henn and J.~C. Plefka,
  \href{http://dx.doi.org/10.1007/978-3-642-54022-6}{{\em {Scattering
  Amplitudes in Gauge Theories}}}, vol.~883.
\newblock Springer, Berlin, 2014.

\bibitem{Sachs:1962wk}
R.~K. Sachs, ``{Gravitational waves in general relativity. 8. Waves in
  asymptotically flat space-times},''
\href{http://dx.doi.org/10.1098/rspa.1962.0206}{{\em Proc. Roy. Soc. Lond.}
  {\bfseries A270} (1962) 103--126}.

\bibitem{Barnich:2010eb}
G.~Barnich and C.~Troessaert, ``{Aspects of the BMS/CFT correspondence},''
  \href{http://dx.doi.org/10.1007/JHEP05(2010)062}{{\em JHEP} {\bfseries 05}
  (2010) 062},
\href{http://arxiv.org/abs/1001.1541}{{\ttfamily arXiv:1001.1541 [hep-th]}}.

\bibitem{Conde:2016rom}
E.~Conde and P.~Mao, ``{BMS Supertranslations and Not So Soft Gravitons},''
  \href{http://dx.doi.org/10.1007/JHEP05(2017)060}{{\em JHEP} {\bfseries 05}
  (2017) 060}, \href{http://arxiv.org/abs/1612.08294}{{\ttfamily
  arXiv:1612.08294 [hep-th]}}.

\bibitem{Lu:2019jus}
H.~L\"u, P.~Mao, and J.-B. Wu, ``{Asymptotic Structure of
  Einstein-Maxwell-Dilaton Theory and Its Five Dimensional Origin},''
  \href{http://dx.doi.org/10.1007/JHEP11(2019)005}{{\em JHEP} {\bfseries 11}
  (2019) 005}, \href{http://arxiv.org/abs/1909.00970}{{\ttfamily
  arXiv:1909.00970 [hep-th]}}.

\bibitem{Ehlers:1962zz}
J.~Ehlers and W.~Kundt, ``{Exact solutions of the gravitational field
  equations},'' in {\em Gravitation: an Introduction to Current Research},
  pp.~49--101.
\newblock John Wiley \& Sons, Inc., New Jersey$\cdot$London, 1962.

\bibitem{Griffiths:2009dfa}
J.~B. Griffiths and J.~Podolsky,
  \href{http://dx.doi.org/10.1017/CBO9780511635397}{{\em {Exact Space-Times in
  Einstein's General Relativity}}}.
\newblock Cambridge University Press, Cambridge, 2009.

\end{thebibliography}
\end{document}